\documentclass[aps, prd, showkeys, twocolumn, superscriptaddress, nofootinbib, floatfix]{revtex4-2}

\usepackage{amsmath}
\usepackage{graphicx}
\usepackage{dcolumn}
\usepackage{bm}

\usepackage[bookmarksnumbered, pdfpagelabels=true, plainpages=false, colorlinks=true, linkcolor=blue, citecolor=blue, urlcolor=blue]{hyperref}

\usepackage{url}
\usepackage{xcolor}
\usepackage[T1]{fontenc}
\usepackage{amssymb}
\usepackage{subcaption}

\begin{document}

\title{Breakdown of Hydrodynamic Universality in Neutron Star Oscillations}

\author{Gabriel S. Denicol}
\email{gsdenicol@id.uff.br}
\affiliation{Instituto de F\'{\i}sica, Universidade Federal Fluminense, Niter\'{o}i, Rio de Janeiro, 24210-346, Brazil.}

\author{Amanda Guerrieri}
\email{amguerrieri@cbpf.br}
\affiliation{CBPF -- Centro Brasileiro de Pesquisas F\'{\i}sicas, 22290-180, Rio de Janeiro, RJ, Brazil.}

\author{João V. M. Muniz}
\email{joaomotta@id.uff.br}
\affiliation{Instituto de F\'{\i}sica, Universidade Federal Fluminense, Niter\'{o}i, Rio de Janeiro, 24210-346, Brazil.}

\author{Gabriel S. Rocha}
\email{gabrielsr@id.uff.br}
\affiliation{Instituto de F\'{\i}sica, Universidade Federal Fluminense, Niter\'{o}i, Rio de Janeiro, 24210-346, Brazil.}

\author{Raissa F. P. Mendes}
\email{rfpmendes@id.uff.br}
\affiliation{Instituto de F\'{\i}sica, Universidade Federal Fluminense, Niter\'{o}i, Rio de Janeiro, 24210-346, Brazil.}
\affiliation{CBPF -- Centro Brasileiro de Pesquisas F\'{\i}sicas, 22290-180, Rio de Janeiro, RJ, Brazil.}

\begin{abstract}
Hydrodynamic universality refers to the property that different formulations of relativistic dissipative hydrodynamics yield identical predictions in the asymptotic long-wavelength regime. We show that neutron-star oscillations need not reach this regime. Because the finite stellar radius limits the accessible wavelengths and causality imposes a lower bound on microscopic relaxation times, the separation between microscopic and macroscopic scales can become insufficient for hydrodynamic universality to emerge. Comparing linear oscillations in relativistic Navier--Stokes and Israel--Stewart theories, we find that realistic bulk viscosities can produce sizeable modifications of the oscillation spectrum, even at the longest wavelengths. 
%We further show that causal relaxation effects qualitatively alter the damping hierarchy of stellar oscillations and, in some cases, allow nonhydrodynamic modes to dominate the dynamics as the star approaches the threshold of gravitational collapse. 
These results identify neutron-star oscillations as a direct probe of microscopic nonequilibrium dynamics beyond the leading hydrodynamic description.
\end{abstract}

\maketitle

\noindent \textit{Introduction}.
The oscillation spectrum of neutron stars is regarded as a valuable probe of the properties of dense matter. Measurements of stellar oscillation modes may provide constraints on the equation of state of strongly interacting matter and, potentially, on transport coefficients and microscopic relaxation processes operating in the stellar interior \cite{Bauswein:2011tp,Takami:2014zpa,Shibata:2005xz,Dietrich:2020eud,Most:2021zvc,Most:2022yhe,Chabanov:2023blf}. As gravitational-wave observations enter the era of third-generation detectors \cite{ET:2025xjr,Evans:2021gyd}, understanding the physical content encoded in stellar oscillation spectra becomes an increasingly important theoretical challenge.

Interpreting neutron-star oscillation spectra requires an accurate description of dissipative hydrodynamics in the stellar interior. Unlike ideal hydrodynamics, relativistic dissipative hydrodynamics can be formulated in different ways, which differ in their treatment of nonequilibrium degrees of freedom, transient dynamics, and higher-order hydrodynamic corrections \cite{Rezzolla:2013dea,Denicol:2021,Rocha:2023ilf,Wagner:2023jgq,BRSSS,Bemfica:2020zjp,Kovtun:2019hdm,Rocha:2022ind}. While these theories may differ substantially away from equilibrium, every relativistic dissipative hydrodynamic theory is required to reproduce the same leading dissipative corrections in the asymptotic hydrodynamic regime, where perturbations vary over sufficiently large length and time scales compared to the underlying microscopic scales. In this limit, the dynamics is governed by conservation laws together with the relativistic Navier--Stokes constitutive relations \cite{landau:59fluid,Eckart:1940te}, so that macroscopic observables are determined solely by equilibrium thermodynamics and the leading transport coefficients. We shall refer to this asymptotic agreement between different relativistic hydrodynamic theories as \textit{hydrodynamic universality}.

The simplest setting in which hydrodynamic universality can be examined is the spectrum of linear perturbations around homogeneous equilibrium states \cite{Hiscock:1987zz,Denicol:2008ha,Denicol:2021,Pu:2009fj,Brito:2020nou,Gavassino:2023mad,Sammet:2023bfo,deBrito:2025jaz,Gavassino:2023odx,Gavassino:2023qwl}. In an infinite medium, translational invariance allows perturbations to be decomposed into plane waves labeled by a continuous wave number $k$, leading to dispersion relations of the form $\omega=\omega(k)$. The asymptotic hydrodynamic regime corresponds to the long-wavelength limit ($k\rightarrow 0$), where \textit{all} theories of relativistic dissipative hydrodynamics are expected to produce identical predictions to leading order in gradients. For example, at leading dissipative order, the sound-mode dispersion relation is \cite{landau:59fluid}
\begin{equation*}
    \omega (k)= \pm c_s {k} - \frac{1}{2} i \Gamma k^2,
\end{equation*}
with $c_s$ being the speed of sound, and $\Gamma$ the attenuation coefficient, determined by the bulk and shear viscosities and by heat conduction. This leading-order dispersion relation is therefore universal, being completely determined by the relativistic Navier--Stokes constitutive relations \cite{landau:59fluid}. At larger wave numbers, however, higher-order hydrodynamic corrections and transient nonequilibrium degrees of freedom become increasingly important and different formulations of relativistic hydrodynamics need no longer coincide.

Hydrodynamic universality is established in homogeneous matter by taking the long-wavelength limit. The key question is whether this limit can actually be realized in a finite star. The finite extent of the system turns the perturbation problem into a boundary-value problem, so that only a discrete set of eigenfrequencies $\{\omega_{lmn}\}$ is allowed \cite{Nollert:1999ji,Kokkotas:1999bd}, and the admissible wave numbers are bounded from below by $k \gtrsim R^{-1}$, with $R$ being the stellar radius. The limit $k\rightarrow 0$, which underlies the notion of hydrodynamic universality, is therefore not physically accessible: the star can only be probed at a fixed, finite gradient scale set by its own size. Whether this is sufficient to reach the asymptotic hydrodynamic regime depends on the separation between microscopic relaxation scales and the characteristic scales of the oscillation.

In this work, we investigate whether neutron-star oscillations necessarily lie within the regime of hydrodynamic universality by comparing predictions of relativistic Navier--Stokes and Israel--Stewart theories \cite{Israel:1979wp}. We show that since causality imposes a lower bound on the microscopic relaxation time, the asymptotic hydrodynamic regime is not trivially realized. 
As a consequence, for sufficiently large viscosities, the spectrum of linear perturbations can differ substantially from the Navier--Stokes prediction even at long wavelengths, leading to qualitatively different oscillatory dynamics. 
In what follows, we adopt $c=G = 1$ unless stated otherwise. \\

\noindent \textit{Transient Dynamics and the Limits of Hydrodynamic Universality.} 
The dynamics of a dissipative neutron star is determined by the coupled evolution of the spacetime geometry and the stellar fluid. The spacetime is described by Einstein's equations, while the fluid obeys local conservation of energy and momentum,
\begin{equation}
G_{\mu\nu}=8\pi T_{\mu\nu}, \qquad \nabla_\mu T^{\mu\nu}=0.
\label{eq:basic_eqs}
\end{equation}
In this work, we restrict our attention to bulk viscous dissipation, neglecting shear stresses and heat conduction, so that the energy-momentum tensor is written as 
\begin{equation}
T^{\mu\nu} = \mathrm{e} u^\mu u^\nu + (\mathrm{p}+\Pi)\Delta^{\mu\nu},
\end{equation}
where $\mathrm{e}$ denotes the energy density, $\mathrm{p}$ the equilibrium pressure, $u^\mu$ the fluid four-velocity, $\Pi$ the bulk viscous pressure, and
$\Delta^{\mu\nu}=g^{\mu\nu}+u^\mu u^\nu$ is the spatial projector orthogonal to the fluid velocity.
Bulk viscosity is expected to arise predominantly from the finite timescale of weak interactions restoring chemical equilibrium during stellar oscillations \cite{Sawyer:1989dp,Haensel:1992zz,Yakovlev:2000jp,Gavassino:2020kwo} and can provide an important source of dissipation over a broad range of neutron-star conditions \cite{Haensel:2002qw,Most:2022yhe,
Ghosh:2025glz,Yang:2023ogo}.

Equations \eqref{eq:basic_eqs} constitute the fundamental dynamical equations of the system. However, by themselves they are not sufficient to determine the evolution of the fluid. As in any hydrodynamic theory, they must be supplemented by an equation of state (EoS) specifying the thermodynamic properties of the fluid, and by evolution equations for the dissipative degrees of freedom.
Throughout this work, we adopt the Israel-Stewart formulation of relativistic dissipative hydrodynamics \cite{Israel:1976tn,Israel:1979wp}, in which the bulk pressure is promoted to an independent dynamical variable satisfying the relaxation equation
\begin{equation}
\tau_\Pi u^\mu\nabla_\mu\Pi+\Pi=-\zeta\theta,
\label{eq:IS}
\end{equation}
where $\theta=\nabla_\mu u^\mu$ is the expansion scalar, $\zeta$ is the bulk viscosity, and $\tau_\Pi$ denotes the bulk relaxation time\footnote{For simplicity, we restrict ourselves to the minimal Israel-Stewart theory and neglect additional higher-order terms that appear in these equations \cite{Israel:1976tn,Israel:1979wp,Denicol:2012cn} --- these terms are not expected to modify our conclusions.}.

The relaxation term $\tau_\Pi u^\mu\nabla_\mu\Pi$ in Eq.~\eqref{eq:IS} is formally of second order in a gradient expansion, whereas the Navier-Stokes constitutive relation $\Pi=-\zeta\theta$ represents the leading, first-order contribution. Consequently, within the asymptotic hydrodynamic regime, where observables are insensitive to higher-order corrections, the oscillation spectrum is expected to depend primarily on the bulk viscosity $\zeta$, with only subleading sensitivity to the bulk relaxation time $\tau_\Pi$. Our strategy is therefore to probe the dependence of the oscillation spectrum on $\tau_\Pi$ --- or, equivalently, the discrepancy between Israel-Stewart and Navier-Stokes predictions --— as a diagnostic of the breakdown of hydrodynamic universality. Before exploring this idea in a specific setting, we first make a few more general remarks.

Unlike relativistic Navier–Stokes, Israel–Stewart theory treats the dissipative currents as independent dynamical degrees of freedom. Consequently, its (linearized) excitation spectrum contains, in addition to the hydrodynamic modes associated with energy-momentum conservation, a transient (nonhydrodynamic) mode with decay time set by $\tau_\Pi$ \cite{Denicol:2021}. This mode describes the relaxation of the bulk viscous pressure toward its Navier–Stokes value. Whenever this relaxation occurs on time scales much shorter than those characterizing the macroscopic fluid motion, the transient mode rapidly decays and the bulk pressure closely follows the constitutive relation $\Pi\simeq-\zeta\theta$. The validity of this approximation is controlled not by the relaxation time itself, but by the dimensionless quantity
$\omega\tau_\Pi$,
where $\omega$ is a characteristic frequency of the fluid motion. The Navier-Stokes regime is therefore recovered asymptotically whenever
\begin{equation} \label{eq:hydrolimit}
\omega\tau_\Pi\ll1,
\end{equation}
in which case the dissipative currents adiabatically follow the evolution of the hydrodynamic variables.

It is tempting to associate the regime \eqref{eq:hydrolimit} with the formal limit $\tau_\Pi \rightarrow 0$. Such an interpretation is, however, misleading. Unlike relativistic Navier-Stokes, Israel-Stewart theory is hyperbolic and possesses a finite set of characteristic propagation velocities. Requiring these characteristic speeds to remain subluminal in the linear regime imposes lower bounds on the relaxation time \cite{Denicol:2008ha,Pu:2009fj},
\begin{equation} \label{eq:causality}
    \tau_\Pi \geq \frac{\zeta}{(1-c_s^2)(\mathrm{e}+\mathrm{p})}.
\end{equation}
Consequently, the formal limit $\tau_\Pi\rightarrow0$ is incompatible with causality. Since the relaxation time cannot be taken to zero, the asymptotic hydrodynamic regime \eqref{eq:hydrolimit} can only be approached when the characteristic frequencies of the fluid dynamics are sufficiently small. In homogeneous systems, this regime can always be reached by considering perturbations with sufficiently long wavelengths. In compact stars, however, the finite stellar size sets a lower bound on the frequencies of stellar oscillation modes.

More concretely, substituting typical neutron-star parameters into Eq.~\eqref{eq:causality} yields
\begin{equation} \label{eq:putting_numbers}
 \omega \tau_\Pi \gtrsim \frac{0.013}{1-c_s^2} \left( \frac{f}{1 \mathrm{kHz}}\right) \left( \frac{\zeta}{10^{30} \mathrm{g/(cm \,s)}} \right) \left( \frac{\rho_\mathrm{sat}c^2}{\mathrm{e}+\mathrm{p}}\right),
\end{equation}
where $\rho_\mathrm{sat} = 2.7 \times 10^{14} \textrm{g/cm}^3$. Hence, sufficiently large sound speeds and bulk viscosities could place the system outside the asymptotic hydrodynamic regime described by \eqref{eq:hydrolimit}. 
In the remainder of this work, we compare Navier–Stokes and causal Israel–Stewart predictions for radial oscillations and show that neutron-star bulk viscosities may indeed drive the system into a regime where hydrodynamic universality is lost. Radial oscillations provide a clean setting in which to isolate this effect, but we note that the relevant parameter, $\omega\tau_\Pi$ in Eq.~\eqref{eq:putting_numbers}, is comparable for non-radial modes \cite{Andersson:1997rn} and post-merger remnants \cite{Dietrich:2020eud}, suggesting that the phenomenon should extend beyond the radial sector.
\\

\noindent \textit{Numerical results.}
We consider linear perturbations around a spherically symmetric neutron star in hydrostatic equilibrium, following Refs.~\cite{Mendes:2025oib,Keeble:2026bzo}. The background spacetime is described by a solution of the Tolman-Oppenheimer-Volkoff (TOV) equations, obtained by solving Einstein's equations for a static perfect fluid. To investigate the oscillation spectrum, we introduce radial perturbations around this equilibrium configuration and linearize Eqs.~\eqref{eq:basic_eqs} and \eqref{eq:IS}. 

In the frequency domain, assuming a harmonic time dependence (i.e., $\propto e^{-i\omega t}$) for the perturbation variables, the problem reduces to a master linear ordinary differential equation \cite{Mendes:2025oib}. The resulting boundary-value problem, subject to regularity at the stellar center and the appropriate surface boundary conditions, admits two discrete families of complex eigenfrequencies, which we denote by $\{ \omega_n^{\textrm{h}}, \omega_n^{\textrm{nh}} \}$. We adopt the convention that $n \in \mathbb{N}$ labels the number of nodes in the radial profile of the Lagrangian displacement ($\xi$), while the superscripts ``h'' and ``nh'' stand for the hydrodynamic and non-hydrodynamic mode families, respectively, the former being continuously connected to the normal modes of a perfect fluid and the latter being absent in relativistic Navier–Stokes theory.

Before presenting our results, we summarize the assumptions adopted for the EoS and transport coefficients.
Since the microscopic relation between $\tau_\Pi$ and $\zeta$ is not well known for neutron-star matter, we adopt the phenomenological parametrization
\begin{equation}
    \tau_\Pi = c_\Pi \frac{\zeta}{(\mathrm{e}+\mathrm{p})c_s^2},
    \label{eq:relaxtime}
\end{equation}
where $c_\Pi$ is a dimensionless parameter (which vanishes in the Navier-Stokes limit). This form preserves the kinetic-theory expectation that the relaxation time is proportional to $\zeta/(\mathrm{e}+\mathrm{p})$ \cite{Denicol:2021}, while the additional factor of $c_s^{-2}$ is motivated by the structure of the causality condition \eqref{eq:causality}, which then reads
\begin{equation}
    c_\Pi \geq \frac{c_s^2}{1-c_s^2}.
    \label{eq:cPi_causal}
\end{equation}

To assess the robustness of our results, we consider two representative combinations of the EoS and bulk-viscosity prescriptions:
\begin{itemize}
    \item[1.] The SLy9 EoS in GPP parametrization \cite{OBoyle:2020qvf} together with
    \begin{equation}
        \zeta = \tau_\zeta (\mathrm{e}+\mathrm{p}), \quad \tau_\zeta = \text{const}.
    \end{equation}
    \item[2.] A $\gamma = 2.75$ polytropic EoS \cite{DePietri:2014mea} together with
    \begin{equation}
        \zeta = \tau_\zeta c_s^2 (\mathrm{e}+\mathrm{p}), \quad \tau_\zeta e^{-\Phi/2} = \text{const},
    \end{equation} 
    where $e^{-\Phi/2}$ is the gravitational redshift factor.
\end{itemize}
While prescription 1 is likely more realistic, prescription 2 offers the technical advantage of eliminating singular points from the integration domain of the frequency-domain equations \cite{Mendes:2025oib}.
In both cases, the magnitude of the bulk viscosity is controlled by the dimensionful parameter $\tau_\zeta$. We vary $\tau_\zeta$ to explore bulk viscosities in the range $ \zeta \sim 10^{29}$--$10^{31}\,\mathrm{g}/(\mathrm{cm\,s})$ which lie within the range of current microscopic estimates for neutron-star matter \cite{Ripley:2023lsq,Yang:2023ogo}. Although such large viscosities are expected primarily for hot matter, finite-temperature effects on radial oscillations were shown to remain small even in this regime \cite{Guerrieri:2026jbm}, justifying our use of a cold EoS.

\begin{figure}[tb]
    \includegraphics[width=1\linewidth]{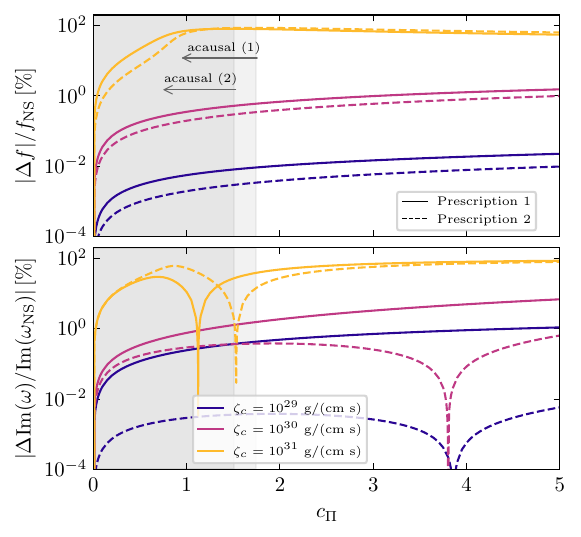}
    \caption{Relative frequency shift of the $n=0$ hydrodynamic mode between Israel-Stewart and Navier-Stokes theories, as a function of $c_\Pi$, for a star with compactness $M/R = 0.25$. Sizable frequency shifts occur at large viscosities.}
    \label{fig:freq_shift}
\end{figure}

\begin{figure}[htb]
    \includegraphics[width=1\linewidth]{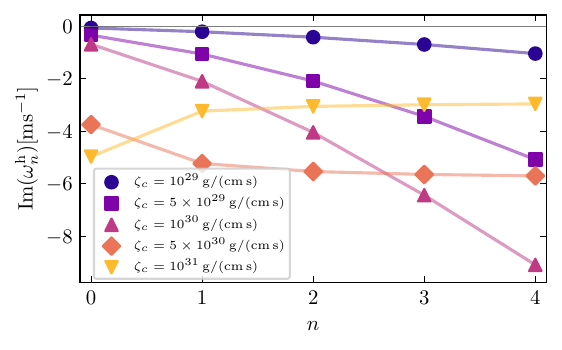}
    \caption{Imaginary part of the hydrodynamic-mode frequencies ($\omega_n^\mathrm{h}$) as a function of the overtone number $n$ for a star with compactness $M/R = 0.25$ and $c_\Pi = 2$ (prescription 2). Symbols denote different central bulk viscosities. At the largest viscosity, the $n=0$ mode is no longer the longest-lived.}
    \label{fig:modes}
\end{figure}

Figure \ref{fig:freq_shift} shows the relative shift in the frequency of the $n=0$ hydrodynamic mode ($\omega =\omega_0^{\textrm{h}}$) between Israel-Stewart and Navier-Stokes theories,
\begin{eqnarray}
2\pi \Delta f &\equiv& \mathrm{Re}(\omega_\mathrm{IS})-\mathrm{Re}(\omega_\mathrm{NS}), \\ \Delta \mathrm{Im}(\omega) &\equiv& \mathrm{Im}(\omega_\mathrm{IS})-\mathrm{Im}(\omega_\mathrm{NS}),   
\end{eqnarray}
as a function of $c_\Pi$, for a star with compactness $M/R = 0.25$.
Different colors denote different values of the central bulk viscosity ($\zeta_c$), while solid (dashed) lines correspond to prescription 1 (2) for the EoS and transport coefficients. Shaded regions correspond to values of $c_\Pi$ that violate the causality condition \eqref{eq:cPi_causal}. 

\begin{figure}[tb]
    \includegraphics[width=1\linewidth]{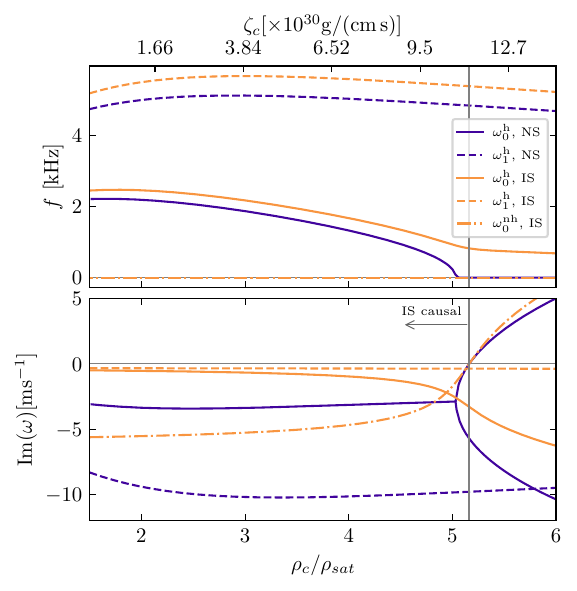}
    \caption{Real (upper panel) and imaginary (lower panel) parts of the radial mode frequencies $\omega^\mathrm{h}_0$ (solid) and $\omega^\mathrm{h}_1$ (dashed) in Navier-Stokes theory (purple), and $\omega^\mathrm{h}_0$ (solid), $\omega^\mathrm{h}_1$ (dashed), and $\omega^{\mathrm{nh}}_0$ (dot-dashed) in Israel-Stewart theory (orange) with $c_\Pi = 11$ (prescription 2), as a function of the central rest-mass density. The bulk-viscosity parameter $\tau_\zeta$ is kept constant along the sequence and chosen such that the maximum-mass neutron star, with central density $\rho_c \approx 5.16\, \rho_\textrm{sat}$ (vertical gray line), has a central bulk viscosity of $\zeta_c = 10^{31}\,\mathrm{g}/(\mathrm{cm\,s})$.}
    \label{fig:freq_density}
\end{figure}

Figure~\ref{fig:freq_shift} shows that, for bulk viscosities of the order $\sim 10^{29}$ g/(cm s), the $n=0$ hydrodynamic modes of both theories are nearly indistinguishable, with $|\Delta f|/f_\textrm{NS} \lesssim 0.01\% $. This illustrates the validity of the asymptotic hydrodynamic regime in the limit of sufficiently weak dissipation. As the bulk viscosity increases, however, the frequency shift becomes progressively more pronounced, reaching $|\Delta f|/f_\textrm{NS} \sim O(10-100)\% $ for $\zeta \sim 10^{31}$ g/(cm s) and causal values for $c_\Pi$. The imaginary part of the eigenfrequencies exhibits a similar behavior to their real part, although $\Delta \mathrm{Im}(\omega)$ changes sign for particular values of $c_\Pi$. 
These results demonstrate that, once causality is imposed and for sufficiently large bulk viscosities, even the longest-wavelength ($n=0$) oscillation mode exhibits a sizeable frequency shift relative to the relativistic Navier–Stokes limit. Therefore, under these conditions, the asymptotic hydrodynamic regime is not reached.
As seen in Fig.~\ref{fig:freq_shift}, these qualitative conclusions seem to be robust under a change in the EoS and bulk-viscosity prescription.

\begin{figure}[tb]
    \includegraphics[width=1\linewidth]{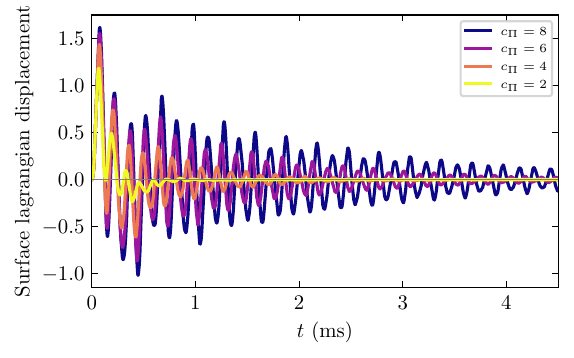}
    \caption{Time evolution of the surface Lagrangian displacement for an initial perturbation $\xi(t=0,r) = \sin(\pi r/R)$ and $\dot\xi(0,r) = \Pi(0,r)=\dot{\Pi}(0,r)= 0$ about a star with $M/R = 0.29$ and $\zeta_c = 10^{31} \mathrm{g/(cm\,s)}$, for $c_\Pi \in \{2,4,6,8\}$ (prescription 1). For this star, the causality condition \eqref{eq:causality} requires $c_\Pi \gtrsim 5.3$. The $c_\Pi = 2$ case, which is closest to the Navier-Stokes limit, already exhibits significant deviations from the causal cases.}
    \label{fig:evolution}
\end{figure}

In the perfect-fluid limit, radial oscillation of relativistic stars form a Sturm–Liouville problem, and the (real) eigenfrequencies obey $(\omega^\textrm{h}_0)^2 < (\omega^\textrm{h}_1)^2 < \cdots$. In the presence of bulk viscosity, however, the governing operator is no longer Hermitian, and no ordering of the eigenfrequencies is guaranteed. In particular, for sufficiently large bulk viscosities, the $n=0$ hydrodynamic mode ($\omega_0^\mathrm{h}$) need not be the longest-lived. As shown in Fig.~\ref{fig:modes}, for a representative star with $M/R = 0.25$ and $\zeta_c = 10^{31} \mathrm{g/(cm\, s)}$, the $n=0$ mode is instead the shortest-lived hydrodynamic mode. 

This inversion is a consequence of finite relaxation effects and is absent in Navier--Stokes theory. Figure~\ref{fig:freq_density} shows the radial mode frequencies $\omega^\mathrm{h}_0$ and $\omega^\mathrm{h}_1$ in Navier-Stokes theory, and $\omega^\mathrm{h}_0, \omega^\mathrm{h}_1$, and $\omega^{\mathrm{nh}}_0$ in Israel-Stewart theory with $c_\Pi = 11$, as a function of central density. For this value of $c_\Pi$, the causality condition \eqref{eq:cPi_causal} is satisfied throughout the stable branch. The bulk-viscosity parameter $\tau_\zeta$ is chosen so that $\zeta_c$ lies in the range $10^{30}$-$10^{31}\,\mathrm{g}/(\mathrm{cm\,s})$ along the sequence. 
In Navier-Stokes theory, the $n=0$ mode remains the longest-lived throughout the sequence. 
In Israel-Stewart theory, by contrast, higher overtones become longer-lived than the $n=0$ mode over part of the stable branch, reversing the usual hierarchy of damping times.

To illustrate the resulting dynamics, Fig.~\ref{fig:evolution} shows the time evolution of an initial perturbation for causal and acausal values of $c_\Pi$. The Navier--Stokes theory itself cannot be evolved directly in time because of its infinite signal propagation speed \cite{hiscock1983stability,hiscock:85generic,pichon:65etude}. Its behavior can nevertheless be approached by considering Israel--Stewart models with $c_\Pi$ below the causality threshold while remaining sufficiently large to avoid numerical instabilities. For the star considered in Fig.~\ref{fig:evolution}, the causality condition requires $c_\Pi\gtrsim5.3$. The evolution for the lowest value considered, $c_\Pi=2$, visibly differs from the causal cases, whose late-time behavior is governed by long-lived, high-frequency modes. 

The spectrum also reveals a second departure from the hydrodynamic picture. In causal Israel-Stewart theory, non-hydrodynamic modes associated with the relaxation dynamics can become dynamically relevant. In particular, for the high-viscosity configurations considered in Fig.~\ref{fig:freq_density}, the onset of the instability at high central densities \cite{Chandrasekhar:1964zza} is governed by the non-hydrodynamic mode $\omega^\mathrm{nh}_0$, rather than by hydrodynamic mode $\omega^\mathrm{h}_0$, as pointed out in Ref.~\cite{Mendes:2025oib}. 

These results demonstrate that for $\zeta \gtrsim  10^{30}\,\mathrm{g}/(\mathrm{cm\,s})$ the characteristic frequencies of stellar oscillations may no longer be well separated from the microscopic relaxation rate, preventing the bulk viscous pressure from adiabatically following the hydrodynamic motion. The spectrum then retains sensitivity to finite relaxation effects, allowing causal corrections to qualitatively alter the stellar oscillation dynamics.\\

\noindent \textit{Conclusions.} 
Neutron-star oscillations need not lie within the regime of hydrodynamic universality if sufficiently large bulk viscosities are realized in dynamical processes involving neutron stars. The finite size of the star limits the accessible wavelengths, while causality imposes a lower bound on microscopic relaxation times, preventing the asymptotic hydrodynamic regime from always being reached.

In the case of radial oscillations, and for bulk viscosities in the upper end of the currently estimated range, $\zeta \sim 10^{30}$--$10^{31} \,\mathrm{g/(cm\,s)}$, we have shown that causal relaxation effects leave sizeable imprints on the spectrum, including substantial shifts in the mode frequencies, a reversal of the damping-time hierarchy among hydrodynamic modes, and the emergence of nonhydrodynamic modes as the dominant contribution, particularly near the maximum-mass configuration. Although our numerical calculations were restricted to radial oscillations, similar effects are expected to arise for non-radial modes \cite{Redondo-Yuste:2024vdb,Katagiri:2026jgp,Bussieres:2026rnz}, since the relevant dimensionless parameter governing causal relaxation, $\omega\tau_\Pi$ in Eq.~\eqref{eq:putting_numbers}, is of comparable magnitude for both classes of oscillations. 
Neutron-star oscillations may thus offer a rare window into how relativistic hydrodynamics emerges from microscopic nonequilibrium dynamics.

We note that our conclusions extend beyond the specific comparison between Israel--Stewart and relativistic Navier--Stokes theories. Whenever neutron-star dynamics fall outside the asymptotic hydrodynamic regime, different formulations of relativistic dissipative hydrodynamics need not produce identical astrophysical predictions, and the choice of framework becomes part of the physical modeling rather than a matter of numerical convenience. This observation is particularly relevant as alternative causal formulations, including BDNK \cite{Bemfica:2017wps, Bemfica:2019knx, Bemfica:2020zjp, Kovtun:2019hdm}, are increasingly employed in numerical-relativity simulations \cite{Pandya:2022pif,Pandya:2022sff,Bantilan:2022ech,Shum:2025jnl,Camelio:2022ljs,Camelio:2022fds,Alford:2017rxf}. The robustness of these simulations ultimately depends on whether the relevant dynamical scales lie within the regime of hydrodynamic universality. 

\noindent \textit{Acknowledgments.} The authors thank Jorge Noronha for fruitful discussions. G.~S.~D.~is supported by CNPq through the grant 307761/2022-3. G.~S.~D.~and G.~S.~R.~acknowledge the support of CNPq and FAPERJ via the INCT-FNA grant 408419/2024-5. A.~G. and J.~M.~acknowledge financial support from CAPES (Coordenação de Aperfeiçoamento de Pessoal de Nível Superior). R.~M.~acknowledges financial support from CNPq, grant 304104/2025-6, as well as FAPERJ, grant E-26/204.589/2024.

\bibliography{refs}

\begin{thebibliography}{63}%
\makeatletter
\providecommand \@ifxundefined [1]{%
 \@ifx{#1\undefined}
}%
\providecommand \@ifnum [1]{%
 \ifnum #1\expandafter \@firstoftwo
 \else \expandafter \@secondoftwo
 \fi
}%
\providecommand \@ifx [1]{%
 \ifx #1\expandafter \@firstoftwo
 \else \expandafter \@secondoftwo
 \fi
}%
\providecommand \natexlab [1]{#1}%
\providecommand \enquote  [1]{``#1''}%
\providecommand \bibnamefont  [1]{#1}%
\providecommand \bibfnamefont [1]{#1}%
\providecommand \citenamefont [1]{#1}%
\providecommand \href@noop [0]{\@secondoftwo}%
\providecommand \href [0]{\begingroup \@sanitize@url \@href}%
\providecommand \@href[1]{\@@startlink{#1}\@@href}%
\providecommand \@@href[1]{\endgroup#1\@@endlink}%
\providecommand \@sanitize@url [0]{\catcode `\\12\catcode `\$12\catcode
  `\&12\catcode `\#12\catcode `\^12\catcode `\_12\catcode `\%12\relax}%
\providecommand \@@startlink[1]{}%
\providecommand \@@endlink[0]{}%
\providecommand \url  [0]{\begingroup\@sanitize@url \@url }%
\providecommand \@url [1]{\endgroup\@href {#1}{\urlprefix }}%
\providecommand \urlprefix  [0]{URL }%
\providecommand \Eprint [0]{\href }%
\providecommand \doibase [0]{https://doi.org/}%
\providecommand \selectlanguage [0]{\@gobble}%
\providecommand \bibinfo  [0]{\@secondoftwo}%
\providecommand \bibfield  [0]{\@secondoftwo}%
\providecommand \translation [1]{[#1]}%
\providecommand \BibitemOpen [0]{}%
\providecommand \bibitemStop [0]{}%
\providecommand \bibitemNoStop [0]{.\EOS\space}%
\providecommand \EOS [0]{\spacefactor3000\relax}%
\providecommand \BibitemShut  [1]{\csname bibitem#1\endcsname}%
\let\auto@bib@innerbib\@empty
%</preamble>
\bibitem [{\citenamefont {Bauswein}\ and\ \citenamefont
  {Janka}(2012)}]{Bauswein:2011tp}%
  \BibitemOpen
  \bibfield  {author} {\bibinfo {author} {\bibfnamefont {A.}~\bibnamefont
  {Bauswein}}\ and\ \bibinfo {author} {\bibfnamefont {H.~T.}\ \bibnamefont
  {Janka}},\ }\href {https://doi.org/10.1103/PhysRevLett.108.011101} {\bibfield
   {journal} {\bibinfo  {journal} {Phys. Rev. Lett.}\ }\textbf {\bibinfo
  {volume} {108}},\ \bibinfo {pages} {011101} (\bibinfo {year} {2012})},\
  \Eprint {https://arxiv.org/abs/1106.1616} {arXiv:1106.1616 [astro-ph.SR]}
  \BibitemShut {NoStop}%
\bibitem [{\citenamefont {Takami}\ \emph {et~al.}(2014)\citenamefont {Takami},
  \citenamefont {Rezzolla},\ and\ \citenamefont {Baiotti}}]{Takami:2014zpa}%
  \BibitemOpen
  \bibfield  {author} {\bibinfo {author} {\bibfnamefont {K.}~\bibnamefont
  {Takami}}, \bibinfo {author} {\bibfnamefont {L.}~\bibnamefont {Rezzolla}},\
  and\ \bibinfo {author} {\bibfnamefont {L.}~\bibnamefont {Baiotti}},\ }\href
  {https://doi.org/10.1103/PhysRevLett.113.091104} {\bibfield  {journal}
  {\bibinfo  {journal} {Phys. Rev. Lett.}\ }\textbf {\bibinfo {volume} {113}},\
  \bibinfo {pages} {091104} (\bibinfo {year} {2014})},\ \Eprint
  {https://arxiv.org/abs/1403.5672} {arXiv:1403.5672 [gr-qc]} \BibitemShut
  {NoStop}%
\bibitem [{\citenamefont {Shibata}(2005)}]{Shibata:2005xz}%
  \BibitemOpen
  \bibfield  {author} {\bibinfo {author} {\bibfnamefont {M.}~\bibnamefont
  {Shibata}},\ }\href {https://doi.org/10.1103/PhysRevLett.94.201101}
  {\bibfield  {journal} {\bibinfo  {journal} {Phys. Rev. Lett.}\ }\textbf
  {\bibinfo {volume} {94}},\ \bibinfo {pages} {201101} (\bibinfo {year}
  {2005})},\ \Eprint {https://arxiv.org/abs/gr-qc/0504082}
  {arXiv:gr-qc/0504082} \BibitemShut {NoStop}%
\bibitem [{\citenamefont {Dietrich}\ \emph {et~al.}(2021)\citenamefont
  {Dietrich}, \citenamefont {Hinderer},\ and\ \citenamefont
  {Samajdar}}]{Dietrich:2020eud}%
  \BibitemOpen
  \bibfield  {author} {\bibinfo {author} {\bibfnamefont {T.}~\bibnamefont
  {Dietrich}}, \bibinfo {author} {\bibfnamefont {T.}~\bibnamefont {Hinderer}},\
  and\ \bibinfo {author} {\bibfnamefont {A.}~\bibnamefont {Samajdar}},\ }\href
  {https://doi.org/10.1007/s10714-020-02751-6} {\bibfield  {journal} {\bibinfo
  {journal} {Gen. Rel. Grav.}\ }\textbf {\bibinfo {volume} {53}},\ \bibinfo
  {pages} {27} (\bibinfo {year} {2021})},\ \Eprint
  {https://arxiv.org/abs/2004.02527} {arXiv:2004.02527 [gr-qc]} \BibitemShut
  {NoStop}%
\bibitem [{\citenamefont {Most}\ \emph {et~al.}(2021)\citenamefont {Most},
  \citenamefont {Harris}, \citenamefont {Plumberg}, \citenamefont {Alford},
  \citenamefont {Noronha}, \citenamefont {Noronha-Hostler}, \citenamefont
  {Pretorius}, \citenamefont {Witek},\ and\ \citenamefont
  {Yunes}}]{Most:2021zvc}%
  \BibitemOpen
  \bibfield  {author} {\bibinfo {author} {\bibfnamefont {E.~R.}\ \bibnamefont
  {Most}}, \bibinfo {author} {\bibfnamefont {S.~P.}\ \bibnamefont {Harris}},
  \bibinfo {author} {\bibfnamefont {C.}~\bibnamefont {Plumberg}}, \bibinfo
  {author} {\bibfnamefont {M.~G.}\ \bibnamefont {Alford}}, \bibinfo {author}
  {\bibfnamefont {J.}~\bibnamefont {Noronha}}, \bibinfo {author} {\bibfnamefont
  {J.}~\bibnamefont {Noronha-Hostler}}, \bibinfo {author} {\bibfnamefont
  {F.}~\bibnamefont {Pretorius}}, \bibinfo {author} {\bibfnamefont
  {H.}~\bibnamefont {Witek}},\ and\ \bibinfo {author} {\bibfnamefont
  {N.}~\bibnamefont {Yunes}},\ }\href {https://doi.org/10.1093/mnras/stab2793}
  {\bibfield  {journal} {\bibinfo  {journal} {Mon. Not. Roy. Astron. Soc.}\
  }\textbf {\bibinfo {volume} {509}},\ \bibinfo {pages} {1096} (\bibinfo {year}
  {2021})},\ \Eprint {https://arxiv.org/abs/2107.05094} {arXiv:2107.05094
  [astro-ph.HE]} \BibitemShut {NoStop}%
\bibitem [{\citenamefont {Most}\ \emph {et~al.}(2024)\citenamefont {Most},
  \citenamefont {Haber}, \citenamefont {Harris}, \citenamefont {Zhang},
  \citenamefont {Alford},\ and\ \citenamefont {Noronha}}]{Most:2022yhe}%
  \BibitemOpen
  \bibfield  {author} {\bibinfo {author} {\bibfnamefont {E.~R.}\ \bibnamefont
  {Most}}, \bibinfo {author} {\bibfnamefont {A.}~\bibnamefont {Haber}},
  \bibinfo {author} {\bibfnamefont {S.~P.}\ \bibnamefont {Harris}}, \bibinfo
  {author} {\bibfnamefont {Z.}~\bibnamefont {Zhang}}, \bibinfo {author}
  {\bibfnamefont {M.~G.}\ \bibnamefont {Alford}},\ and\ \bibinfo {author}
  {\bibfnamefont {J.}~\bibnamefont {Noronha}},\ }\href
  {https://doi.org/10.3847/2041-8213/ad454f} {\bibfield  {journal} {\bibinfo
  {journal} {Astrophys. J. Lett.}\ }\textbf {\bibinfo {volume} {967}},\
  \bibinfo {pages} {L14} (\bibinfo {year} {2024})},\ \Eprint
  {https://arxiv.org/abs/2207.00442} {arXiv:2207.00442 [astro-ph.HE]}
  \BibitemShut {NoStop}%
\bibitem [{\citenamefont {Chabanov}\ and\ \citenamefont
  {Rezzolla}(2025)}]{Chabanov:2023blf}%
  \BibitemOpen
  \bibfield  {author} {\bibinfo {author} {\bibfnamefont {M.}~\bibnamefont
  {Chabanov}}\ and\ \bibinfo {author} {\bibfnamefont {L.}~\bibnamefont
  {Rezzolla}},\ }\href {https://doi.org/10.1103/PhysRevLett.134.071402}
  {\bibfield  {journal} {\bibinfo  {journal} {Phys. Rev. Lett.}\ }\textbf
  {\bibinfo {volume} {134}},\ \bibinfo {pages} {071402} (\bibinfo {year}
  {2025})},\ \Eprint {https://arxiv.org/abs/2307.10464} {arXiv:2307.10464
  [gr-qc]} \BibitemShut {NoStop}%
\bibitem [{\citenamefont {Abac}\ \emph {et~al.}(2026)\citenamefont {Abac} \emph
  {et~al.}}]{ET:2025xjr}%
  \BibitemOpen
  \bibfield  {author} {\bibinfo {author} {\bibfnamefont {A.}~\bibnamefont
  {Abac}} \emph {et~al.} (\bibinfo {collaboration} {ET}),\ }\href
  {https://doi.org/10.1088/1475-7516/2026/03/081} {\bibfield  {journal}
  {\bibinfo  {journal} {JCAP}\ }\textbf {\bibinfo {volume} {03}},\ \bibinfo
  {pages} {081}},\ \Eprint {https://arxiv.org/abs/2503.12263} {arXiv:2503.12263
  [gr-qc]} \BibitemShut {NoStop}%
\bibitem [{\citenamefont {Evans}\ \emph {et~al.}(2021)\citenamefont {Evans}
  \emph {et~al.}}]{Evans:2021gyd}%
  \BibitemOpen
  \bibfield  {author} {\bibinfo {author} {\bibfnamefont {M.}~\bibnamefont
  {Evans}} \emph {et~al.},\ }\href@noop {} {\bibinfo {title} {{A Horizon Study
  for Cosmic Explorer: Science, Observatories, and Community}}} (\bibinfo
  {year} {2021}),\ \Eprint {https://arxiv.org/abs/2109.09882} {arXiv:2109.09882
  [astro-ph.IM]} \BibitemShut {NoStop}%
\bibitem [{\citenamefont {Rezzolla}\ and\ \citenamefont
  {Zanotti}(2013)}]{Rezzolla:2013dea}%
  \BibitemOpen
  \bibfield  {author} {\bibinfo {author} {\bibfnamefont {L.}~\bibnamefont
  {Rezzolla}}\ and\ \bibinfo {author} {\bibfnamefont {O.}~\bibnamefont
  {Zanotti}},\ }\href
  {https://doi.org/10.1093/acprof:oso/9780198528906.001.0001} {\emph {\bibinfo
  {title} {{Relativistic Hydrodynamics}}}}\ (\bibinfo  {publisher} {Oxford
  University Press},\ \bibinfo {year} {2013})\BibitemShut {NoStop}%
\bibitem [{\citenamefont {Denicol}\ and\ \citenamefont
  {Rischke}(2021)}]{Denicol:2021}%
  \BibitemOpen
  \bibfield  {author} {\bibinfo {author} {\bibfnamefont {G.~S.}\ \bibnamefont
  {Denicol}}\ and\ \bibinfo {author} {\bibfnamefont {D.~H.}\ \bibnamefont
  {Rischke}},\ }\href@noop {} {\emph {\bibinfo {title} {Microscopic Foundations
  of Relativistic Fluid Dynamics}}}\ (\bibinfo  {publisher} {Springer},\
  \bibinfo {year} {2021})\BibitemShut {NoStop}%
\bibitem [{\citenamefont {Rocha}\ \emph {et~al.}(2024)\citenamefont {Rocha},
  \citenamefont {Wagner}, \citenamefont {Denicol}, \citenamefont {Noronha},\
  and\ \citenamefont {Rischke}}]{Rocha:2023ilf}%
  \BibitemOpen
  \bibfield  {author} {\bibinfo {author} {\bibfnamefont {G.~S.}\ \bibnamefont
  {Rocha}}, \bibinfo {author} {\bibfnamefont {D.}~\bibnamefont {Wagner}},
  \bibinfo {author} {\bibfnamefont {G.~S.}\ \bibnamefont {Denicol}}, \bibinfo
  {author} {\bibfnamefont {J.}~\bibnamefont {Noronha}},\ and\ \bibinfo {author}
  {\bibfnamefont {D.~H.}\ \bibnamefont {Rischke}},\ }\href
  {https://doi.org/10.3390/e26030189} {\bibfield  {journal} {\bibinfo
  {journal} {Entropy}\ }\textbf {\bibinfo {volume} {26}},\ \bibinfo {pages}
  {189} (\bibinfo {year} {2024})},\ \Eprint {https://arxiv.org/abs/2311.15063}
  {arXiv:2311.15063 [nucl-th]} \BibitemShut {NoStop}%
\bibitem [{\citenamefont {Wagner}\ and\ \citenamefont
  {Gavassino}(2024)}]{Wagner:2023jgq}%
  \BibitemOpen
  \bibfield  {author} {\bibinfo {author} {\bibfnamefont {D.}~\bibnamefont
  {Wagner}}\ and\ \bibinfo {author} {\bibfnamefont {L.}~\bibnamefont
  {Gavassino}},\ }\href {https://doi.org/10.1103/PhysRevD.109.016019}
  {\bibfield  {journal} {\bibinfo  {journal} {Phys. Rev. D}\ }\textbf {\bibinfo
  {volume} {109}},\ \bibinfo {pages} {016019} (\bibinfo {year} {2024})},\
  \Eprint {https://arxiv.org/abs/2309.14828} {arXiv:2309.14828 [nucl-th]}
  \BibitemShut {NoStop}%
\bibitem [{\citenamefont {Baier}\ \emph {et~al.}(2008)\citenamefont {Baier},
  \citenamefont {Romatschke}, \citenamefont {Son}, \citenamefont {Starinets},\
  and\ \citenamefont {Stephanov}}]{BRSSS}%
  \BibitemOpen
  \bibfield  {author} {\bibinfo {author} {\bibfnamefont {R.}~\bibnamefont
  {Baier}}, \bibinfo {author} {\bibfnamefont {P.}~\bibnamefont {Romatschke}},
  \bibinfo {author} {\bibfnamefont {D.~T.}\ \bibnamefont {Son}}, \bibinfo
  {author} {\bibfnamefont {A.~O.}\ \bibnamefont {Starinets}},\ and\ \bibinfo
  {author} {\bibfnamefont {M.~A.}\ \bibnamefont {Stephanov}},\ }\href
  {https://doi.org/10.1088/1126-6708/2008/04/100} {\bibfield  {journal}
  {\bibinfo  {journal} {JHEP}\ }\textbf {\bibinfo {volume} {2008}}\bibinfo
  {number} { (04)},\ \bibinfo {pages} {100–100}}\BibitemShut {NoStop}%
\bibitem [{\citenamefont {Bemfica}\ \emph {et~al.}(2022)\citenamefont
  {Bemfica}, \citenamefont {Disconzi},\ and\ \citenamefont
  {Noronha}}]{Bemfica:2020zjp}%
  \BibitemOpen
\bibfield  {number} {  }\bibfield  {author} {\bibinfo {author} {\bibfnamefont
  {F.~S.}\ \bibnamefont {Bemfica}}, \bibinfo {author} {\bibfnamefont {M.~M.}\
  \bibnamefont {Disconzi}},\ and\ \bibinfo {author} {\bibfnamefont
  {J.}~\bibnamefont {Noronha}},\ }\href
  {https://doi.org/10.1103/PhysRevX.12.021044} {\bibfield  {journal} {\bibinfo
  {journal} {Phys. Rev. X}\ }\textbf {\bibinfo {volume} {12}},\ \bibinfo
  {pages} {021044} (\bibinfo {year} {2022})},\ \Eprint
  {https://arxiv.org/abs/2009.11388} {arXiv:2009.11388 [gr-qc]} \BibitemShut
  {NoStop}%
\bibitem [{\citenamefont {Kovtun}(2019)}]{Kovtun:2019hdm}%
  \BibitemOpen
  \bibfield  {author} {\bibinfo {author} {\bibfnamefont {P.}~\bibnamefont
  {Kovtun}},\ }\href {https://doi.org/10.1007/JHEP10(2019)034} {\bibfield
  {journal} {\bibinfo  {journal} {JHEP}\ }\textbf {\bibinfo {volume} {10}},\
  \bibinfo {pages} {034}},\ \Eprint {https://arxiv.org/abs/1907.08191}
  {arXiv:1907.08191 [hep-th]} \BibitemShut {NoStop}%
\bibitem [{\citenamefont {Rocha}\ \emph {et~al.}(2022)\citenamefont {Rocha},
  \citenamefont {Denicol},\ and\ \citenamefont {Noronha}}]{Rocha:2022ind}%
  \BibitemOpen
  \bibfield  {author} {\bibinfo {author} {\bibfnamefont {G.~S.}\ \bibnamefont
  {Rocha}}, \bibinfo {author} {\bibfnamefont {G.~S.}\ \bibnamefont {Denicol}},\
  and\ \bibinfo {author} {\bibfnamefont {J.}~\bibnamefont {Noronha}},\ }\href
  {https://doi.org/10.1103/PhysRevD.106.036010} {\bibfield  {journal} {\bibinfo
   {journal} {Phys. Rev. D}\ }\textbf {\bibinfo {volume} {106}},\ \bibinfo
  {pages} {036010} (\bibinfo {year} {2022})},\ \Eprint
  {https://arxiv.org/abs/2205.00078} {arXiv:2205.00078 [nucl-th]} \BibitemShut
  {NoStop}%
\bibitem [{\citenamefont {Landau}\ and\ \citenamefont
  {Lifshitz}(1959)}]{landau:59fluid}%
  \BibitemOpen
  \bibfield  {author} {\bibinfo {author} {\bibfnamefont {L.}~\bibnamefont
  {Landau}}\ and\ \bibinfo {author} {\bibfnamefont {E.}~\bibnamefont
  {Lifshitz}},\ }\href@noop {} {\emph {\bibinfo {title} {Fluid {M}echanics}}},\
  \bibinfo {series} {Course of Theoretical Physics}, Vol.~\bibinfo {volume}
  {6}\ (\bibinfo  {publisher} {Addison-Wesley},\ \bibinfo {year}
  {1959})\BibitemShut {NoStop}%
\bibitem [{\citenamefont {Eckart}(1940)}]{Eckart:1940te}%
  \BibitemOpen
  \bibfield  {author} {\bibinfo {author} {\bibfnamefont {C.}~\bibnamefont
  {Eckart}},\ }\href {https://doi.org/10.1103/PhysRev.58.919} {\bibfield
  {journal} {\bibinfo  {journal} {Phys. Rev.}\ }\textbf {\bibinfo {volume}
  {58}},\ \bibinfo {pages} {919} (\bibinfo {year} {1940})}\BibitemShut
  {NoStop}%
\bibitem [{\citenamefont {Hiscock}\ and\ \citenamefont
  {Lindblom}(1987)}]{Hiscock:1987zz}%
  \BibitemOpen
  \bibfield  {author} {\bibinfo {author} {\bibfnamefont {W.~A.}\ \bibnamefont
  {Hiscock}}\ and\ \bibinfo {author} {\bibfnamefont {L.}~\bibnamefont
  {Lindblom}},\ }\href {https://doi.org/10.1103/PhysRevD.35.3723} {\bibfield
  {journal} {\bibinfo  {journal} {Phys. Rev. D}\ }\textbf {\bibinfo {volume}
  {35}},\ \bibinfo {pages} {3723} (\bibinfo {year} {1987})}\BibitemShut
  {NoStop}%
\bibitem [{\citenamefont {Denicol}\ \emph {et~al.}(2008)\citenamefont
  {Denicol}, \citenamefont {Kodama}, \citenamefont {Koide},\ and\ \citenamefont
  {Mota}}]{Denicol:2008ha}%
  \BibitemOpen
  \bibfield  {author} {\bibinfo {author} {\bibfnamefont {G.~S.}\ \bibnamefont
  {Denicol}}, \bibinfo {author} {\bibfnamefont {T.}~\bibnamefont {Kodama}},
  \bibinfo {author} {\bibfnamefont {T.}~\bibnamefont {Koide}},\ and\ \bibinfo
  {author} {\bibfnamefont {P.}~\bibnamefont {Mota}},\ }\href
  {https://doi.org/10.1088/0954-3899/35/11/115102} {\bibfield  {journal}
  {\bibinfo  {journal} {J. Phys. G}\ }\textbf {\bibinfo {volume} {35}},\
  \bibinfo {pages} {115102} (\bibinfo {year} {2008})},\ \Eprint
  {https://arxiv.org/abs/0807.3120} {arXiv:0807.3120 [hep-ph]} \BibitemShut
  {NoStop}%
\bibitem [{\citenamefont {Pu}\ \emph {et~al.}(2010)\citenamefont {Pu},
  \citenamefont {Koide},\ and\ \citenamefont {Rischke}}]{Pu:2009fj}%
  \BibitemOpen
  \bibfield  {author} {\bibinfo {author} {\bibfnamefont {S.}~\bibnamefont
  {Pu}}, \bibinfo {author} {\bibfnamefont {T.}~\bibnamefont {Koide}},\ and\
  \bibinfo {author} {\bibfnamefont {D.~H.}\ \bibnamefont {Rischke}},\ }\href
  {https://doi.org/10.1103/PhysRevD.81.114039} {\bibfield  {journal} {\bibinfo
  {journal} {Phys. Rev. D}\ }\textbf {\bibinfo {volume} {81}},\ \bibinfo
  {pages} {114039} (\bibinfo {year} {2010})},\ \Eprint
  {https://arxiv.org/abs/0907.3906} {arXiv:0907.3906 [hep-ph]} \BibitemShut
  {NoStop}%
\bibitem [{\citenamefont {Brito}\ and\ \citenamefont
  {Denicol}(2020)}]{Brito:2020nou}%
  \BibitemOpen
  \bibfield  {author} {\bibinfo {author} {\bibfnamefont {C.~V.}\ \bibnamefont
  {Brito}}\ and\ \bibinfo {author} {\bibfnamefont {G.~S.}\ \bibnamefont
  {Denicol}},\ }\href {https://doi.org/10.1103/PhysRevD.102.116009} {\bibfield
  {journal} {\bibinfo  {journal} {Phys. Rev. D}\ }\textbf {\bibinfo {volume}
  {102}},\ \bibinfo {pages} {116009} (\bibinfo {year} {2020})},\ \Eprint
  {https://arxiv.org/abs/2007.16141} {arXiv:2007.16141 [nucl-th]} \BibitemShut
  {NoStop}%
\bibitem [{\citenamefont {Gavassino}\ \emph
  {et~al.}(2024{\natexlab{a}})\citenamefont {Gavassino}, \citenamefont
  {Disconzi},\ and\ \citenamefont {Noronha}}]{Gavassino:2023mad}%
  \BibitemOpen
  \bibfield  {author} {\bibinfo {author} {\bibfnamefont {L.}~\bibnamefont
  {Gavassino}}, \bibinfo {author} {\bibfnamefont {M.~M.}\ \bibnamefont
  {Disconzi}},\ and\ \bibinfo {author} {\bibfnamefont {J.}~\bibnamefont
  {Noronha}},\ }\href {https://doi.org/10.1103/PhysRevLett.132.162301}
  {\bibfield  {journal} {\bibinfo  {journal} {Phys. Rev. Lett.}\ }\textbf
  {\bibinfo {volume} {132}},\ \bibinfo {pages} {162301} (\bibinfo {year}
  {2024}{\natexlab{a}})},\ \Eprint {https://arxiv.org/abs/2307.05987}
  {arXiv:2307.05987 [hep-th]} \BibitemShut {NoStop}%
\bibitem [{\citenamefont {Sammet}\ \emph {et~al.}(2023)\citenamefont {Sammet},
  \citenamefont {Mayer},\ and\ \citenamefont {Rischke}}]{Sammet:2023bfo}%
  \BibitemOpen
  \bibfield  {author} {\bibinfo {author} {\bibfnamefont {J.}~\bibnamefont
  {Sammet}}, \bibinfo {author} {\bibfnamefont {M.}~\bibnamefont {Mayer}},\ and\
  \bibinfo {author} {\bibfnamefont {D.~H.}\ \bibnamefont {Rischke}},\ }\href
  {https://doi.org/10.1103/PhysRevD.107.114028} {\bibfield  {journal} {\bibinfo
   {journal} {Phys. Rev. D}\ }\textbf {\bibinfo {volume} {107}},\ \bibinfo
  {pages} {114028} (\bibinfo {year} {2023})},\ \Eprint
  {https://arxiv.org/abs/2302.01070} {arXiv:2302.01070 [hep-th]} \BibitemShut
  {NoStop}%
\bibitem [{\citenamefont {de~Brito}\ \emph {et~al.}(2025)\citenamefont
  {de~Brito}, \citenamefont {Kushwah},\ and\ \citenamefont
  {Denicol}}]{deBrito:2025jaz}%
  \BibitemOpen
  \bibfield  {author} {\bibinfo {author} {\bibfnamefont {C.~V.~P.}\
  \bibnamefont {de~Brito}}, \bibinfo {author} {\bibfnamefont {K.}~\bibnamefont
  {Kushwah}},\ and\ \bibinfo {author} {\bibfnamefont {G.~S.}\ \bibnamefont
  {Denicol}},\ }\href {https://doi.org/10.1103/nf7r-z6zk} {\bibfield  {journal}
  {\bibinfo  {journal} {Phys. Rev. D}\ }\textbf {\bibinfo {volume} {112}},\
  \bibinfo {pages} {076035} (\bibinfo {year} {2025})},\ \Eprint
  {https://arxiv.org/abs/2505.10397} {arXiv:2505.10397 [nucl-th]} \BibitemShut
  {NoStop}%
\bibitem [{\citenamefont {Gavassino}\ \emph
  {et~al.}(2024{\natexlab{b}})\citenamefont {Gavassino}, \citenamefont
  {Disconzi},\ and\ \citenamefont {Noronha}}]{Gavassino:2023odx}%
  \BibitemOpen
  \bibfield  {author} {\bibinfo {author} {\bibfnamefont {L.}~\bibnamefont
  {Gavassino}}, \bibinfo {author} {\bibfnamefont {M.~M.}\ \bibnamefont
  {Disconzi}},\ and\ \bibinfo {author} {\bibfnamefont {J.}~\bibnamefont
  {Noronha}},\ }\href {https://doi.org/10.1103/PhysRevLett.132.222302}
  {\bibfield  {journal} {\bibinfo  {journal} {Phys. Rev. Lett.}\ }\textbf
  {\bibinfo {volume} {132}},\ \bibinfo {pages} {222302} (\bibinfo {year}
  {2024}{\natexlab{b}})},\ \Eprint {https://arxiv.org/abs/2302.03478}
  {arXiv:2302.03478 [nucl-th]} \BibitemShut {NoStop}%
\bibitem [{\citenamefont {Gavassino}\ \emph
  {et~al.}(2024{\natexlab{c}})\citenamefont {Gavassino}, \citenamefont
  {Disconzi},\ and\ \citenamefont {Noronha}}]{Gavassino:2023qwl}%
  \BibitemOpen
  \bibfield  {author} {\bibinfo {author} {\bibfnamefont {L.}~\bibnamefont
  {Gavassino}}, \bibinfo {author} {\bibfnamefont {M.~M.}\ \bibnamefont
  {Disconzi}},\ and\ \bibinfo {author} {\bibfnamefont {J.}~\bibnamefont
  {Noronha}},\ }\href {https://doi.org/10.1103/PhysRevD.109.096041} {\bibfield
  {journal} {\bibinfo  {journal} {Phys. Rev. D}\ }\textbf {\bibinfo {volume}
  {109}},\ \bibinfo {pages} {096041} (\bibinfo {year} {2024}{\natexlab{c}})},\
  \Eprint {https://arxiv.org/abs/2302.05332} {arXiv:2302.05332 [nucl-th]}
  \BibitemShut {NoStop}%
\bibitem [{\citenamefont {Nollert}(1999)}]{Nollert:1999ji}%
  \BibitemOpen
  \bibfield  {author} {\bibinfo {author} {\bibfnamefont {H.-P.}\ \bibnamefont
  {Nollert}},\ }\href {https://doi.org/10.1088/0264-9381/16/12/201} {\bibfield
  {journal} {\bibinfo  {journal} {Class. Quant. Grav.}\ }\textbf {\bibinfo
  {volume} {16}},\ \bibinfo {pages} {R159} (\bibinfo {year}
  {1999})}\BibitemShut {NoStop}%
\bibitem [{\citenamefont {Kokkotas}\ and\ \citenamefont
  {Schmidt}(1999)}]{Kokkotas:1999bd}%
  \BibitemOpen
  \bibfield  {author} {\bibinfo {author} {\bibfnamefont {K.~D.}\ \bibnamefont
  {Kokkotas}}\ and\ \bibinfo {author} {\bibfnamefont {B.~G.}\ \bibnamefont
  {Schmidt}},\ }\href {https://doi.org/10.12942/lrr-1999-2} {\bibfield
  {journal} {\bibinfo  {journal} {Living Rev. Rel.}\ }\textbf {\bibinfo
  {volume} {2}},\ \bibinfo {pages} {2} (\bibinfo {year} {1999})},\ \Eprint
  {https://arxiv.org/abs/gr-qc/9909058} {arXiv:gr-qc/9909058} \BibitemShut
  {NoStop}%
\bibitem [{\citenamefont {Israel}\ and\ \citenamefont
  {Stewart}(1979)}]{Israel:1979wp}%
  \BibitemOpen
  \bibfield  {author} {\bibinfo {author} {\bibfnamefont {W.}~\bibnamefont
  {Israel}}\ and\ \bibinfo {author} {\bibfnamefont {J.~M.}\ \bibnamefont
  {Stewart}},\ }\href {https://doi.org/10.1016/0003-4916(79)90130-1} {\bibfield
   {journal} {\bibinfo  {journal} {Annals Phys.}\ }\textbf {\bibinfo {volume}
  {118}},\ \bibinfo {pages} {341} (\bibinfo {year} {1979})}\BibitemShut
  {NoStop}%
\bibitem [{\citenamefont {Sawyer}(1989)}]{Sawyer:1989dp}%
  \BibitemOpen
  \bibfield  {author} {\bibinfo {author} {\bibfnamefont {R.~F.}\ \bibnamefont
  {Sawyer}},\ }\href {https://doi.org/10.1103/PhysRevD.39.3804} {\bibfield
  {journal} {\bibinfo  {journal} {Phys. Rev. D}\ }\textbf {\bibinfo {volume}
  {39}},\ \bibinfo {pages} {3804} (\bibinfo {year} {1989})}\BibitemShut
  {NoStop}%
\bibitem [{\citenamefont {Haensel}\ and\ \citenamefont
  {Schaeffer}(1992)}]{Haensel:1992zz}%
  \BibitemOpen
  \bibfield  {author} {\bibinfo {author} {\bibfnamefont {P.}~\bibnamefont
  {Haensel}}\ and\ \bibinfo {author} {\bibfnamefont {R.}~\bibnamefont
  {Schaeffer}},\ }\href {https://doi.org/10.1103/PhysRevD.45.4708} {\bibfield
  {journal} {\bibinfo  {journal} {Phys. Rev. D}\ }\textbf {\bibinfo {volume}
  {45}},\ \bibinfo {pages} {4708} (\bibinfo {year} {1992})}\BibitemShut
  {NoStop}%
\bibitem [{\citenamefont {Yakovlev}\ \emph {et~al.}(2001)\citenamefont
  {Yakovlev}, \citenamefont {Kaminker}, \citenamefont {Gnedin},\ and\
  \citenamefont {Haensel}}]{Yakovlev:2000jp}%
  \BibitemOpen
  \bibfield  {author} {\bibinfo {author} {\bibfnamefont {D.~G.}\ \bibnamefont
  {Yakovlev}}, \bibinfo {author} {\bibfnamefont {A.~D.}\ \bibnamefont
  {Kaminker}}, \bibinfo {author} {\bibfnamefont {O.~Y.}\ \bibnamefont
  {Gnedin}},\ and\ \bibinfo {author} {\bibfnamefont {P.}~\bibnamefont
  {Haensel}},\ }\href {https://doi.org/10.1016/S0370-1573(00)00131-9}
  {\bibfield  {journal} {\bibinfo  {journal} {Phys. Rept.}\ }\textbf {\bibinfo
  {volume} {354}},\ \bibinfo {pages} {1} (\bibinfo {year} {2001})},\ \Eprint
  {https://arxiv.org/abs/astro-ph/0012122} {arXiv:astro-ph/0012122}
  \BibitemShut {NoStop}%
\bibitem [{\citenamefont {Gavassino}\ \emph {et~al.}(2021)\citenamefont
  {Gavassino}, \citenamefont {Antonelli},\ and\ \citenamefont
  {Haskell}}]{Gavassino:2020kwo}%
  \BibitemOpen
  \bibfield  {author} {\bibinfo {author} {\bibfnamefont {L.}~\bibnamefont
  {Gavassino}}, \bibinfo {author} {\bibfnamefont {M.}~\bibnamefont
  {Antonelli}},\ and\ \bibinfo {author} {\bibfnamefont {B.}~\bibnamefont
  {Haskell}},\ }\href {https://doi.org/10.1088/1361-6382/abe588} {\bibfield
  {journal} {\bibinfo  {journal} {Class. Quant. Grav.}\ }\textbf {\bibinfo
  {volume} {38}},\ \bibinfo {pages} {075001} (\bibinfo {year} {2021})},\
  \Eprint {https://arxiv.org/abs/2003.04609} {arXiv:2003.04609 [gr-qc]}
  \BibitemShut {NoStop}%
\bibitem [{\citenamefont {Haensel}\ \emph {et~al.}(2002)\citenamefont
  {Haensel}, \citenamefont {Levenfish},\ and\ \citenamefont
  {Yakovlev}}]{Haensel:2002qw}%
  \BibitemOpen
  \bibfield  {author} {\bibinfo {author} {\bibfnamefont {P.}~\bibnamefont
  {Haensel}}, \bibinfo {author} {\bibfnamefont {K.~P.}\ \bibnamefont
  {Levenfish}},\ and\ \bibinfo {author} {\bibfnamefont {D.~G.}\ \bibnamefont
  {Yakovlev}},\ }\href {https://doi.org/10.1051/0004-6361:20021112} {\bibfield
  {journal} {\bibinfo  {journal} {Astron. Astrophys.}\ }\textbf {\bibinfo
  {volume} {394}},\ \bibinfo {pages} {213} (\bibinfo {year} {2002})},\ \Eprint
  {https://arxiv.org/abs/astro-ph/0208078} {arXiv:astro-ph/0208078}
  \BibitemShut {NoStop}%
\bibitem [{\citenamefont {Ghosh}\ \emph {et~al.}(2025)\citenamefont {Ghosh},
  \citenamefont {Mukherjee}, \citenamefont {Bose},\ and\ \citenamefont
  {Chatterjee}}]{Ghosh:2025glz}%
  \BibitemOpen
  \bibfield  {author} {\bibinfo {author} {\bibfnamefont {S.}~\bibnamefont
  {Ghosh}}, \bibinfo {author} {\bibfnamefont {S.}~\bibnamefont {Mukherjee}},
  \bibinfo {author} {\bibfnamefont {S.}~\bibnamefont {Bose}},\ and\ \bibinfo
  {author} {\bibfnamefont {D.}~\bibnamefont {Chatterjee}},\ }\href
  {https://doi.org/10.1093/mnras/staf1652} {\bibfield  {journal} {\bibinfo
  {journal} {Mon. Not. Roy. Astron. Soc.}\ }\textbf {\bibinfo {volume} {543}},\
  \bibinfo {pages} {2987} (\bibinfo {year} {2025})},\ \Eprint
  {https://arxiv.org/abs/2503.14606} {arXiv:2503.14606 [gr-qc]} \BibitemShut
  {NoStop}%
\bibitem [{\citenamefont {Yang}\ \emph {et~al.}(2024)\citenamefont {Yang},
  \citenamefont {Hippert}, \citenamefont {Speranza},\ and\ \citenamefont
  {Noronha}}]{Yang:2023ogo}%
  \BibitemOpen
  \bibfield  {author} {\bibinfo {author} {\bibfnamefont {Y.}~\bibnamefont
  {Yang}}, \bibinfo {author} {\bibfnamefont {M.}~\bibnamefont {Hippert}},
  \bibinfo {author} {\bibfnamefont {E.}~\bibnamefont {Speranza}},\ and\
  \bibinfo {author} {\bibfnamefont {J.}~\bibnamefont {Noronha}},\ }\href
  {https://doi.org/10.1103/PhysRevC.109.015805} {\bibfield  {journal} {\bibinfo
   {journal} {Phys. Rev. C}\ }\textbf {\bibinfo {volume} {109}},\ \bibinfo
  {pages} {015805} (\bibinfo {year} {2024})},\ \Eprint
  {https://arxiv.org/abs/2309.01864} {arXiv:2309.01864 [nucl-th]} \BibitemShut
  {NoStop}%
\bibitem [{\citenamefont {Israel}(1976)}]{Israel:1976tn}%
  \BibitemOpen
  \bibfield  {author} {\bibinfo {author} {\bibfnamefont {W.}~\bibnamefont
  {Israel}},\ }\href {https://doi.org/10.1016/0003-4916(76)90064-6} {\bibfield
  {journal} {\bibinfo  {journal} {Annals Phys.}\ }\textbf {\bibinfo {volume}
  {100}},\ \bibinfo {pages} {310} (\bibinfo {year} {1976})}\BibitemShut
  {NoStop}%
\bibitem [{\citenamefont {Denicol}\ \emph {et~al.}(2012)\citenamefont
  {Denicol}, \citenamefont {Niemi}, \citenamefont {Molnar},\ and\ \citenamefont
  {Rischke}}]{Denicol:2012cn}%
  \BibitemOpen
  \bibfield  {author} {\bibinfo {author} {\bibfnamefont {G.~S.}\ \bibnamefont
  {Denicol}}, \bibinfo {author} {\bibfnamefont {H.}~\bibnamefont {Niemi}},
  \bibinfo {author} {\bibfnamefont {E.}~\bibnamefont {Molnar}},\ and\ \bibinfo
  {author} {\bibfnamefont {D.~H.}\ \bibnamefont {Rischke}},\ }\href
  {https://doi.org/10.1103/PhysRevD.85.114047} {\bibfield  {journal} {\bibinfo
  {journal} {Phys. Rev. D}\ }\textbf {\bibinfo {volume} {85}},\ \bibinfo
  {pages} {114047} (\bibinfo {year} {2012})},\ \bibinfo {note} {[Erratum:
  Phys.Rev.D 91, 039902 (2015)]},\ \Eprint {https://arxiv.org/abs/1202.4551}
  {arXiv:1202.4551 [nucl-th]} \BibitemShut {NoStop}%
\bibitem [{\citenamefont {Andersson}\ and\ \citenamefont
  {Kokkotas}(1998)}]{Andersson:1997rn}%
  \BibitemOpen
  \bibfield  {author} {\bibinfo {author} {\bibfnamefont {N.}~\bibnamefont
  {Andersson}}\ and\ \bibinfo {author} {\bibfnamefont {K.~D.}\ \bibnamefont
  {Kokkotas}},\ }\href {https://doi.org/10.1046/j.1365-8711.1998.01840.x}
  {\bibfield  {journal} {\bibinfo  {journal} {Mon. Not. Roy. Astron. Soc.}\
  }\textbf {\bibinfo {volume} {299}},\ \bibinfo {pages} {1059} (\bibinfo {year}
  {1998})},\ \Eprint {https://arxiv.org/abs/gr-qc/9711088}
  {arXiv:gr-qc/9711088} \BibitemShut {NoStop}%
\bibitem [{\citenamefont {Mendes}\ \emph {et~al.}(2026)\citenamefont {Mendes},
  \citenamefont {Guerrieri}, \citenamefont {Muniz}, \citenamefont {Rocha},\
  and\ \citenamefont {Denicol}}]{Mendes:2025oib}%
  \BibitemOpen
  \bibfield  {author} {\bibinfo {author} {\bibfnamefont {R.~F.~P.}\
  \bibnamefont {Mendes}}, \bibinfo {author} {\bibfnamefont {A.}~\bibnamefont
  {Guerrieri}}, \bibinfo {author} {\bibfnamefont {J.~V.~M.}\ \bibnamefont
  {Muniz}}, \bibinfo {author} {\bibfnamefont {G.~S.}\ \bibnamefont {Rocha}},\
  and\ \bibinfo {author} {\bibfnamefont {G.~S.}\ \bibnamefont {Denicol}},\
  }\href {https://doi.org/10.1103/fs15-pj7m} {\bibfield  {journal} {\bibinfo
  {journal} {Phys. Rev. D}\ }\textbf {\bibinfo {volume} {113}},\ \bibinfo
  {pages} {024010} (\bibinfo {year} {2026})},\ \Eprint
  {https://arxiv.org/abs/2509.12330} {arXiv:2509.12330 [gr-qc]} \BibitemShut
  {NoStop}%
\bibitem [{\citenamefont {Keeble}\ and\ \citenamefont
  {Redondo-Yuste}(2026)}]{Keeble:2026bzo}%
  \BibitemOpen
  \bibfield  {author} {\bibinfo {author} {\bibfnamefont {L.~S.}\ \bibnamefont
  {Keeble}}\ and\ \bibinfo {author} {\bibfnamefont {J.}~\bibnamefont
  {Redondo-Yuste}},\ }\href@noop {} {\bibinfo {title} {{Radial Oscillations of
  Viscous Stars}}} (\bibinfo {year} {2026}),\ \Eprint
  {https://arxiv.org/abs/2603.23622} {arXiv:2603.23622 [gr-qc]} \BibitemShut
  {NoStop}%
\bibitem [{\citenamefont {O'Boyle}\ \emph {et~al.}(2020)\citenamefont
  {O'Boyle}, \citenamefont {Markakis}, \citenamefont {Stergioulas},\ and\
  \citenamefont {Read}}]{OBoyle:2020qvf}%
  \BibitemOpen
  \bibfield  {author} {\bibinfo {author} {\bibfnamefont {M.~F.}\ \bibnamefont
  {O'Boyle}}, \bibinfo {author} {\bibfnamefont {C.}~\bibnamefont {Markakis}},
  \bibinfo {author} {\bibfnamefont {N.}~\bibnamefont {Stergioulas}},\ and\
  \bibinfo {author} {\bibfnamefont {J.~S.}\ \bibnamefont {Read}},\ }\href
  {https://doi.org/10.1103/PhysRevD.102.083027} {\bibfield  {journal} {\bibinfo
   {journal} {Phys. Rev. D}\ }\textbf {\bibinfo {volume} {102}},\ \bibinfo
  {pages} {083027} (\bibinfo {year} {2020})},\ \Eprint
  {https://arxiv.org/abs/2008.03342} {arXiv:2008.03342 [astro-ph.HE]}
  \BibitemShut {NoStop}%
\bibitem [{\citenamefont {De~Pietri}\ \emph {et~al.}(2014)\citenamefont
  {De~Pietri}, \citenamefont {Feo}, \citenamefont {Franci},\ and\ \citenamefont
  {L{\"o}ffler}}]{DePietri:2014mea}%
  \BibitemOpen
  \bibfield  {author} {\bibinfo {author} {\bibfnamefont {R.}~\bibnamefont
  {De~Pietri}}, \bibinfo {author} {\bibfnamefont {A.}~\bibnamefont {Feo}},
  \bibinfo {author} {\bibfnamefont {L.}~\bibnamefont {Franci}},\ and\ \bibinfo
  {author} {\bibfnamefont {F.}~\bibnamefont {L{\"o}ffler}},\ }\href
  {https://doi.org/10.1103/PhysRevD.90.024034} {\bibfield  {journal} {\bibinfo
  {journal} {Phys. Rev. D}\ }\textbf {\bibinfo {volume} {90}},\ \bibinfo
  {pages} {024034} (\bibinfo {year} {2014})},\ \Eprint
  {https://arxiv.org/abs/1403.8066} {arXiv:1403.8066 [gr-qc]} \BibitemShut
  {NoStop}%
\bibitem [{\citenamefont {Ripley}\ \emph {et~al.}(2024)\citenamefont {Ripley},
  \citenamefont {Hegade K.~R.}, \citenamefont {Chandramouli},\ and\
  \citenamefont {Yunes}}]{Ripley:2023lsq}%
  \BibitemOpen
  \bibfield  {author} {\bibinfo {author} {\bibfnamefont {J.~L.}\ \bibnamefont
  {Ripley}}, \bibinfo {author} {\bibfnamefont {A.}~\bibnamefont {Hegade
  K.~R.}}, \bibinfo {author} {\bibfnamefont {R.~S.}\ \bibnamefont
  {Chandramouli}},\ and\ \bibinfo {author} {\bibfnamefont {N.}~\bibnamefont
  {Yunes}},\ }\href {https://doi.org/10.1038/s41550-024-02323-7} {\bibfield
  {journal} {\bibinfo  {journal} {Nature Astron.}\ }\textbf {\bibinfo {volume}
  {8}},\ \bibinfo {pages} {1277} (\bibinfo {year} {2024})},\ \Eprint
  {https://arxiv.org/abs/2312.11659} {arXiv:2312.11659 [gr-qc]} \BibitemShut
  {NoStop}%
\bibitem [{\citenamefont {Guerrieri}\ \emph {et~al.}(2026)\citenamefont
  {Guerrieri}, \citenamefont {Rocha}, \citenamefont {Denicol},\ and\
  \citenamefont {Mendes}}]{Guerrieri:2026jbm}%
  \BibitemOpen
  \bibfield  {author} {\bibinfo {author} {\bibfnamefont {A.}~\bibnamefont
  {Guerrieri}}, \bibinfo {author} {\bibfnamefont {G.~S.}\ \bibnamefont
  {Rocha}}, \bibinfo {author} {\bibfnamefont {G.~S.}\ \bibnamefont {Denicol}},\
  and\ \bibinfo {author} {\bibfnamefont {R.~F.~P.}\ \bibnamefont {Mendes}},\
  }\href@noop {} {\bibinfo {title} {{Radial Oscillations of Viscous Stars at
  Finite Temperature}}} (\bibinfo {year} {2026}),\ \Eprint
  {https://arxiv.org/abs/2606.06600} {arXiv:2606.06600 [gr-qc]} \BibitemShut
  {NoStop}%
\bibitem [{\citenamefont {Hiscock}\ and\ \citenamefont
  {Lindblom}(1983)}]{hiscock1983stability}%
  \BibitemOpen
  \bibfield  {author} {\bibinfo {author} {\bibfnamefont {W.~A.}\ \bibnamefont
  {Hiscock}}\ and\ \bibinfo {author} {\bibfnamefont {L.}~\bibnamefont
  {Lindblom}},\ }\href@noop {} {\bibfield  {journal} {\bibinfo  {journal}
  {Annals of Physics}\ }\textbf {\bibinfo {volume} {151}},\ \bibinfo {pages}
  {466} (\bibinfo {year} {1983})}\BibitemShut {NoStop}%
\bibitem [{\citenamefont {Hiscock}\ and\ \citenamefont
  {Lindblom}(1985)}]{hiscock:85generic}%
  \BibitemOpen
  \bibfield  {author} {\bibinfo {author} {\bibfnamefont {W.~A.}\ \bibnamefont
  {Hiscock}}\ and\ \bibinfo {author} {\bibfnamefont {L.}~\bibnamefont
  {Lindblom}},\ }\href@noop {} {\bibfield  {journal} {\bibinfo  {journal}
  {Physical Review D}\ }\textbf {\bibinfo {volume} {31}},\ \bibinfo {pages}
  {725} (\bibinfo {year} {1985})}\BibitemShut {NoStop}%
\bibitem [{\citenamefont {Pichon}(1965)}]{pichon:65etude}%
  \BibitemOpen
  \bibfield  {author} {\bibinfo {author} {\bibfnamefont {G.}~\bibnamefont
  {Pichon}},\ }in\ \href@noop {} {\emph {\bibinfo {booktitle} {Annales de l'IHP
  Physique th{\'e}orique}}},\ Vol.~\bibinfo {volume} {2}\ (\bibinfo {year}
  {1965})\ pp.\ \bibinfo {pages} {21--85}\BibitemShut {NoStop}%
\bibitem [{\citenamefont {Chandrasekhar}(1964)}]{Chandrasekhar:1964zza}%
  \BibitemOpen
  \bibfield  {author} {\bibinfo {author} {\bibfnamefont {S.}~\bibnamefont
  {Chandrasekhar}},\ }\href {https://doi.org/10.1103/PhysRevLett.12.114}
  {\bibfield  {journal} {\bibinfo  {journal} {Phys. Rev. Lett.}\ }\textbf
  {\bibinfo {volume} {12}},\ \bibinfo {pages} {114} (\bibinfo {year}
  {1964})}\BibitemShut {NoStop}%
\bibitem [{\citenamefont {Redondo-Yuste}(2025)}]{Redondo-Yuste:2024vdb}%
  \BibitemOpen
  \bibfield  {author} {\bibinfo {author} {\bibfnamefont {J.}~\bibnamefont
  {Redondo-Yuste}},\ }\href {https://doi.org/10.1088/1361-6382/adbfef}
  {\bibfield  {journal} {\bibinfo  {journal} {Class. Quant. Grav.}\ }\textbf
  {\bibinfo {volume} {42}},\ \bibinfo {pages} {075012} (\bibinfo {year}
  {2025})},\ \Eprint {https://arxiv.org/abs/2411.16841} {arXiv:2411.16841
  [gr-qc]} \BibitemShut {NoStop}%
\bibitem [{\citenamefont {Katagiri}\ \emph {et~al.}(2026)\citenamefont
  {Katagiri}, \citenamefont {Guedes},\ and\ \citenamefont
  {Yagi}}]{Katagiri:2026jgp}%
  \BibitemOpen
  \bibfield  {author} {\bibinfo {author} {\bibfnamefont {T.}~\bibnamefont
  {Katagiri}}, \bibinfo {author} {\bibfnamefont {V.}~\bibnamefont {Guedes}},\
  and\ \bibinfo {author} {\bibfnamefont {K.}~\bibnamefont {Yagi}},\ }\href@noop
  {} {\bibinfo {title} {{Out-of-Equilibrium Effects in Non-Radial Relativistic
  Stellar Perturbations: A Model-Agnostic Formulation and Mode Analysis}}}
  (\bibinfo {year} {2026}),\ \Eprint {https://arxiv.org/abs/2606.20957}
  {arXiv:2606.20957 [gr-qc]} \BibitemShut {NoStop}%
\bibitem [{\citenamefont {Bussi{\`e}res}\ \emph {et~al.}(2026)\citenamefont
  {Bussi{\`e}res}, \citenamefont {Redondo-Yuste}, \citenamefont
  {Ortega~G{\'o}mez},\ and\ \citenamefont {Cardoso}}]{Bussieres:2026rnz}%
  \BibitemOpen
  \bibfield  {author} {\bibinfo {author} {\bibfnamefont {S.}~\bibnamefont
  {Bussi{\`e}res}}, \bibinfo {author} {\bibfnamefont {J.}~\bibnamefont
  {Redondo-Yuste}}, \bibinfo {author} {\bibfnamefont {J.~J.}\ \bibnamefont
  {Ortega~G{\'o}mez}},\ and\ \bibinfo {author} {\bibfnamefont {V.}~\bibnamefont
  {Cardoso}},\ }\href@noop {} {\bibinfo {title} {{Axial Oscillations of Viscous
  Neutron Stars}}} (\bibinfo {year} {2026}),\ \Eprint
  {https://arxiv.org/abs/2604.13208} {arXiv:2604.13208 [gr-qc]} \BibitemShut
  {NoStop}%
\bibitem [{\citenamefont {Bemfica}\ \emph {et~al.}(2018)\citenamefont
  {Bemfica}, \citenamefont {Disconzi},\ and\ \citenamefont
  {Noronha}}]{Bemfica:2017wps}%
  \BibitemOpen
  \bibfield  {author} {\bibinfo {author} {\bibfnamefont {F.}~\bibnamefont
  {Bemfica}}, \bibinfo {author} {\bibfnamefont {M.}~\bibnamefont {Disconzi}},\
  and\ \bibinfo {author} {\bibfnamefont {J.}~\bibnamefont {Noronha}},\ }\href
  {https://doi.org/10.1103/PhysRevD.98.104064} {\bibfield  {journal} {\bibinfo
  {journal} {Phys. Rev. D}\ }\textbf {\bibinfo {volume} {98}},\ \bibinfo
  {pages} {104064} (\bibinfo {year} {2018})},\ \Eprint
  {https://arxiv.org/abs/1708.06255} {arXiv:1708.06255 [gr-qc]} \BibitemShut
  {NoStop}%
\bibitem [{\citenamefont {Bemfica}\ \emph {et~al.}(2019)\citenamefont
  {Bemfica}, \citenamefont {Disconzi},\ and\ \citenamefont
  {Noronha}}]{Bemfica:2019knx}%
  \BibitemOpen
  \bibfield  {author} {\bibinfo {author} {\bibfnamefont {F.~S.}\ \bibnamefont
  {Bemfica}}, \bibinfo {author} {\bibfnamefont {M.~M.}\ \bibnamefont
  {Disconzi}},\ and\ \bibinfo {author} {\bibfnamefont {J.}~\bibnamefont
  {Noronha}},\ }\href {https://doi.org/10.1103/PhysRevD.100.104020} {\bibfield
  {journal} {\bibinfo  {journal} {Phys. Rev. D}\ }\textbf {\bibinfo {volume}
  {100}},\ \bibinfo {pages} {104020} (\bibinfo {year} {2019})},\ \bibinfo
  {note} {[Erratum: Phys.Rev.D 105, 069902 (2022)]},\ \Eprint
  {https://arxiv.org/abs/1907.12695} {arXiv:1907.12695 [gr-qc]} \BibitemShut
  {NoStop}%
\bibitem [{\citenamefont {Pandya}\ \emph
  {et~al.}(2022{\natexlab{a}})\citenamefont {Pandya}, \citenamefont {Most},\
  and\ \citenamefont {Pretorius}}]{Pandya:2022pif}%
  \BibitemOpen
  \bibfield  {author} {\bibinfo {author} {\bibfnamefont {A.}~\bibnamefont
  {Pandya}}, \bibinfo {author} {\bibfnamefont {E.~R.}\ \bibnamefont {Most}},\
  and\ \bibinfo {author} {\bibfnamefont {F.}~\bibnamefont {Pretorius}},\ }\href
  {https://doi.org/10.1103/PhysRevD.105.123001} {\bibfield  {journal} {\bibinfo
   {journal} {Phys. Rev. D}\ }\textbf {\bibinfo {volume} {105}},\ \bibinfo
  {pages} {123001} (\bibinfo {year} {2022}{\natexlab{a}})},\ \Eprint
  {https://arxiv.org/abs/2201.12317} {arXiv:2201.12317 [gr-qc]} \BibitemShut
  {NoStop}%
\bibitem [{\citenamefont {Pandya}\ \emph
  {et~al.}(2022{\natexlab{b}})\citenamefont {Pandya}, \citenamefont {Most},\
  and\ \citenamefont {Pretorius}}]{Pandya:2022sff}%
  \BibitemOpen
  \bibfield  {author} {\bibinfo {author} {\bibfnamefont {A.}~\bibnamefont
  {Pandya}}, \bibinfo {author} {\bibfnamefont {E.~R.}\ \bibnamefont {Most}},\
  and\ \bibinfo {author} {\bibfnamefont {F.}~\bibnamefont {Pretorius}},\ }\href
  {https://doi.org/10.1103/PhysRevD.106.123036} {\bibfield  {journal} {\bibinfo
   {journal} {Phys. Rev. D}\ }\textbf {\bibinfo {volume} {106}},\ \bibinfo
  {pages} {123036} (\bibinfo {year} {2022}{\natexlab{b}})},\ \Eprint
  {https://arxiv.org/abs/2209.09265} {arXiv:2209.09265 [gr-qc]} \BibitemShut
  {NoStop}%
\bibitem [{\citenamefont {Bantilan}\ \emph {et~al.}(2022)\citenamefont
  {Bantilan}, \citenamefont {Bea},\ and\ \citenamefont
  {Figueras}}]{Bantilan:2022ech}%
  \BibitemOpen
  \bibfield  {author} {\bibinfo {author} {\bibfnamefont {H.}~\bibnamefont
  {Bantilan}}, \bibinfo {author} {\bibfnamefont {Y.}~\bibnamefont {Bea}},\ and\
  \bibinfo {author} {\bibfnamefont {P.}~\bibnamefont {Figueras}},\ }\href
  {https://doi.org/10.1007/JHEP08(2022)298} {\bibfield  {journal} {\bibinfo
  {journal} {JHEP}\ }\textbf {\bibinfo {volume} {08}},\ \bibinfo {pages}
  {298}},\ \Eprint {https://arxiv.org/abs/2201.13359} {arXiv:2201.13359
  [hep-th]} \BibitemShut {NoStop}%
\bibitem [{\citenamefont {Shum}\ \emph {et~al.}(2026)\citenamefont {Shum},
  \citenamefont {Abalos}, \citenamefont {Bea}, \citenamefont {Bezares},
  \citenamefont {Figueras},\ and\ \citenamefont {Palenzuela}}]{Shum:2025jnl}%
  \BibitemOpen
  \bibfield  {author} {\bibinfo {author} {\bibfnamefont {H.~L.~H.}\
  \bibnamefont {Shum}}, \bibinfo {author} {\bibfnamefont {F.}~\bibnamefont
  {Abalos}}, \bibinfo {author} {\bibfnamefont {Y.}~\bibnamefont {Bea}},
  \bibinfo {author} {\bibfnamefont {M.}~\bibnamefont {Bezares}}, \bibinfo
  {author} {\bibfnamefont {P.}~\bibnamefont {Figueras}},\ and\ \bibinfo
  {author} {\bibfnamefont {C.}~\bibnamefont {Palenzuela}},\ }\href
  {https://doi.org/10.1103/mthy-4119} {\bibfield  {journal} {\bibinfo
  {journal} {Phys. Rev. D}\ }\textbf {\bibinfo {volume} {113}},\ \bibinfo
  {pages} {084029} (\bibinfo {year} {2026})},\ \Eprint
  {https://arxiv.org/abs/2509.15303} {arXiv:2509.15303 [gr-qc]} \BibitemShut
  {NoStop}%
\bibitem [{\citenamefont {Camelio}\ \emph
  {et~al.}(2023{\natexlab{a}})\citenamefont {Camelio}, \citenamefont
  {Gavassino}, \citenamefont {Antonelli}, \citenamefont {Bernuzzi},\ and\
  \citenamefont {Haskell}}]{Camelio:2022ljs}%
  \BibitemOpen
  \bibfield  {author} {\bibinfo {author} {\bibfnamefont {G.}~\bibnamefont
  {Camelio}}, \bibinfo {author} {\bibfnamefont {L.}~\bibnamefont {Gavassino}},
  \bibinfo {author} {\bibfnamefont {M.}~\bibnamefont {Antonelli}}, \bibinfo
  {author} {\bibfnamefont {S.}~\bibnamefont {Bernuzzi}},\ and\ \bibinfo
  {author} {\bibfnamefont {B.}~\bibnamefont {Haskell}},\ }\href
  {https://doi.org/10.1103/PhysRevD.107.103031} {\bibfield  {journal} {\bibinfo
   {journal} {Phys. Rev. D}\ }\textbf {\bibinfo {volume} {107}},\ \bibinfo
  {pages} {103031} (\bibinfo {year} {2023}{\natexlab{a}})},\ \Eprint
  {https://arxiv.org/abs/2204.11809} {arXiv:2204.11809 [gr-qc]} \BibitemShut
  {NoStop}%
\bibitem [{\citenamefont {Camelio}\ \emph
  {et~al.}(2023{\natexlab{b}})\citenamefont {Camelio}, \citenamefont
  {Gavassino}, \citenamefont {Antonelli}, \citenamefont {Bernuzzi},\ and\
  \citenamefont {Haskell}}]{Camelio:2022fds}%
  \BibitemOpen
  \bibfield  {author} {\bibinfo {author} {\bibfnamefont {G.}~\bibnamefont
  {Camelio}}, \bibinfo {author} {\bibfnamefont {L.}~\bibnamefont {Gavassino}},
  \bibinfo {author} {\bibfnamefont {M.}~\bibnamefont {Antonelli}}, \bibinfo
  {author} {\bibfnamefont {S.}~\bibnamefont {Bernuzzi}},\ and\ \bibinfo
  {author} {\bibfnamefont {B.}~\bibnamefont {Haskell}},\ }\href
  {https://doi.org/10.1103/PhysRevD.107.103032} {\bibfield  {journal} {\bibinfo
   {journal} {Phys. Rev. D}\ }\textbf {\bibinfo {volume} {107}},\ \bibinfo
  {pages} {103032} (\bibinfo {year} {2023}{\natexlab{b}})},\ \Eprint
  {https://arxiv.org/abs/2204.11810} {arXiv:2204.11810 [gr-qc]} \BibitemShut
  {NoStop}%
\bibitem [{\citenamefont {Alford}\ \emph {et~al.}(2018)\citenamefont {Alford},
  \citenamefont {Bovard}, \citenamefont {Hanauske}, \citenamefont {Rezzolla},\
  and\ \citenamefont {Schwenzer}}]{Alford:2017rxf}%
  \BibitemOpen
  \bibfield  {author} {\bibinfo {author} {\bibfnamefont {M.~G.}\ \bibnamefont
  {Alford}}, \bibinfo {author} {\bibfnamefont {L.}~\bibnamefont {Bovard}},
  \bibinfo {author} {\bibfnamefont {M.}~\bibnamefont {Hanauske}}, \bibinfo
  {author} {\bibfnamefont {L.}~\bibnamefont {Rezzolla}},\ and\ \bibinfo
  {author} {\bibfnamefont {K.}~\bibnamefont {Schwenzer}},\ }\href
  {https://doi.org/10.1103/PhysRevLett.120.041101} {\bibfield  {journal}
  {\bibinfo  {journal} {Phys. Rev. Lett.}\ }\textbf {\bibinfo {volume} {120}},\
  \bibinfo {pages} {041101} (\bibinfo {year} {2018})},\ \Eprint
  {https://arxiv.org/abs/1707.09475} {arXiv:1707.09475 [gr-qc]} \BibitemShut
  {NoStop}%
\end{thebibliography}%
\bibliographystyle{apsrev4-2}

\end{document}